\documentclass[conference]{IEEEtran}
\IEEEoverridecommandlockouts
\usepackage{cite}
\usepackage{amsmath,amssymb,amsfonts}
\usepackage{algorithmic}
\usepackage{graphicx}
\usepackage{textcomp}
\usepackage{xcolor}
\usepackage{soul}
\usepackage{caption}
\usepackage{subcaption}
\def\BibTeX{{\rm B\kern-.05em{\sc i\kern-.025em b}\kern-.08em
    T\kern-.1667em\lower.7ex\hbox{E}\kern-.125emX}}
\newtheorem{remark}{Remark}
\begin{document}

\title{Image Restoration under Semantic Communications \vspace{-0.2cm} 
\thanks{
This research is funded by Hanoi University of Science and Technology (HUST) under project number T2021-SAHEP-002.}
}

\author{\IEEEauthorblockN{Trinh Van Chien$^\dagger$, Le Hong Phong$^\ast$, Dao Xuan Phuc$^\ddagger$ and Nguyen Tien Hoa$^\ast$}
\IEEEauthorblockA{
	$^\dagger$ School of Information and Communication Technology (SoICT), Hanoi University of Science and Technology, Vietnam\\
	$^\ast$ School of Electronic and Electrical Engineering, Hanoi University of Science and Technology, Vietnam\\
	$^\ddagger$ Faculty of Electrical and Electronic Engineering, Hanoi Open University, Vietnam\\
	}
	Corresponding author: Nguyen Tien Hoa (Email:  hoa.nguyentien@hust.edu.vn.)
}

\maketitle

\begin{abstract}
Semantic communication has emerged as the breakthrough beyond the Shannon theorem by transmitting and receiving semantic information instead of data bits or symbols regardless of its content. This paper proposes a two-stage reconstruction process to boost the system's performance. In the first phase, the image information is first decoded from the noisy received data by exploiting the channel knowledge. The decoded image is enhanced by a post-filter and image statistics. Different metrics are exploited to evaluate the image restoration quality of our considered model. Numerical results are obtained using natural images that verify the superior improvements of the proposed two-stage reconstruction process over the traditional decoded data. Moreover, the different metrics assessing the system performance based on their criteria can be conflicted with each other. 
\end{abstract}

\vspace{-0.2cm}
\section{Introduction}
\vspace{-0.2cm}
Communication technologies have redefined the way people communicate, which have recently attracted a lot of interest via the so-called semantic communications. It has become one of the main engines for propelling the development of society \cite{nguyen2020performance}. Until the development of fifth-generation (5G) mobile networks, most communication systems are designed based on rate metrics, for instance, throughput, spectrum/energy efficiency, and latency. The semantic aspect is treated as irrelevant to an engineering problem by Claude Shannon, and this perspective is still being kept until now \cite{1948-CommunicationTheory}. Yet this point of view is about to change, as we are approaching the Shannon physical capacity limit with advanced encoding (decoding) and modulation techniques, semantic communications are expected to play an important role in sixth-generation \cite{2021-02-6G-Commnunications,2022-02-Semantic2}. In the 5G systems, Massive MIMO (multiple-input multiple-output) technology has become one of the core technologies and it will also be the foundation for the next generations of mobile networks \cite{nguyen2021pilot}. This technology allows focusing the signal to the user by using antenna arrays to beamforming towards the user directly, avoiding energy waste and increasing transmission efficiency\cite{2017-01-MassiveMIMO}.

The transmission of images and videos has been one of the main problems of telecommunications networks for a long time. With the exponential growth rate of data traffic on transferring images, streaming, video conferencing or online learning, which takes up a large number of system resources, requiring longer transmission and processing time, a highly efficient communication paradigm is needed more than ever. Furthermore, image and video data structures are strongly correlated and require a completely different processing mechanism than other random data \cite{sun2020semantic}. As mobile data speeds increase, the amount of multimedia data transmitted will also increase in the following years.
\vspace{-0.2cm}
\subsection{Related Works}
\vspace{-0.1cm}
Many researches in the literature have been carried out to find ways to apply semantic communication to enhance the transmission performance of these multimedia signals. In \cite{21-10-Semantic1}, the authors discussed  the semantic communication principles, architecture, its application areas, and the approach to it with machine learning (ML) and information of things (IoTs). P. Jiang et al.\cite{2022-04-Video-Conference} have proposed semantic-based video conferencing network, where the authors used the incremental redundancy hybrid automatic repeat request (IR-HARQ) feedback framework to guarantee the quality of the transmission according to signal-to-noise ratio (SNR) and channel state information (CSI). A deep learning enabled semantic communication systems (DeepSC-S) for speech signals transmission was proposed in \cite{2021-09-Speech-Transmission}, which learns and extracts speech signals at the transmitter and then recovering them at the receiver from the received features directly. Then, the authors compared its performance with the traditional approaches to verify the effectiveness of the system, especially in  deeply faded environments. To the best of our knowledge, there is no related work considering using semantic information at the receiver to enhance the reconstruction quality and evaluating the system performance by the different evaluation metrics.
\vspace{-0.2cm}
\subsection{Contributions}
\vspace{-0.1cm}
This paper studies an uplink massive multiple-input multiple-output (MIMO) semantic transmission system where a user transmits image data to the base station (BS). The sematic image data includes the spatial correlation between the pixels instead of the randomly non-structure signals.  The data is then modulated and transmitted over the wireless environments. At the BS, the alternate direction method of multipliers (ADMM) is first used to reconstruct the transmitted image signals by minimizing the transmission error over the random fluctuations of  fading channels. After that, we consider using a post-filter to improve the reconstruction quality from the output of the ADMM via mitigating  residual errors based on  image correlation structure \cite{van2013edge}. This paper considers the traditional filters comprising a median filter, a Gaussian filter, and a block-matching and 3D filtering. Finally, the performance of image data recovery is assessed by three parameters commonly used in wireless communication (the physical layer) and multimedia (the application layer), namely, the bit error rate (BER), the peak signal-to-noise ratio (PSNR), and the structural similarity (SSIM) over  different signal-to-noise ratio (SNR) values. Numerical results demonstrate the significant importance in exploiting the prior image information to enhance the reconstruction quality. Besides, the performance metric usually used in the physical layer may be contradicted to one in the application layer.
\vspace{-0.2cm}
\section{System Model and Problem Formulation}
\vspace{-0.1cm}
The semantic communication system involves  two stages, comprising the decoded stage and the post-processing  restoration stage. Specifically, we consider a user communicates with the BS, both equipped with multiple antennas. After receiving the transmitted signals, the image data is decoded at the BS. In the post-transmission restoration phase, the reconstructed image is enhanced by a low-pass filter exploiting image structure.
\begin{figure}
	\centering
	\includegraphics[trim=8cm 12.5cm 13cm 2cm, clip=true, width=2.8in]{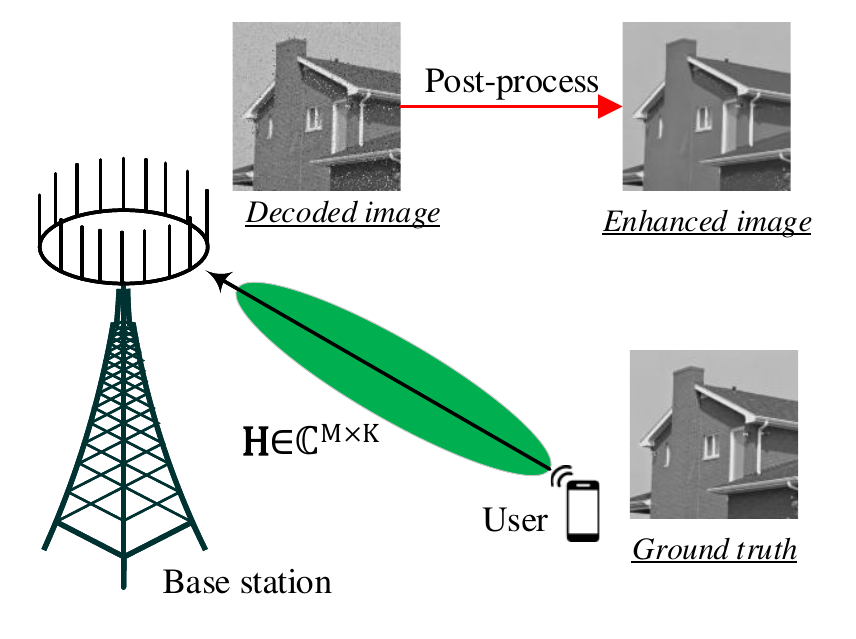}
	\caption{The considered system model where image data content is transmitted over the wireless network.}
	\label{fig:system_model}
	\vspace{-0.5cm}
\end{figure}
\vspace{-0.2cm}
\subsection{System Model}
\vspace{-0.1cm}
We consider an uplink of a Massive MIMO system where the BS is equipped with $M$ antennas communicates with a user having $K$ antennas. Let us suppose that this user selects the semantic content of the  image data $\mathbf{I}$ and sends it to the BS.  The image content will be simply stacked into $Z$ column vectors as follows
    $\mathbf{I} \rightarrow \{\mathbf{i}_1,\cdots,\mathbf{i}_Z\}$. 
These column vectors are further modulated so that each vector is mapped to the constellation points of the alphabet set $\mathcal{O}$. Each value in the vector $\mathbf{i}_z, \forall z = 1, \ldots, Z,$ corresponds to one or several constellation points. This mapping process will be represented  as
   $ \mathbf{I} \overset{\mathcal{M}}{\mapsto} \mathbf{X}  \Leftrightarrow 
    \{i_1,...,i_N\} \mapsto \{x_1,...,x_n\}$, 
where $x_z \in \mathbb{C}, \forall z,$ are the modulated symbols that will be transmitted over the MIMO channel and received at the BS; $\mathcal{M}$ indicates the mapping or the modulation. In particular,  let us define $\mathbf{y}_z$ the received signal at the BS after propagating over the environment as
\begin{equation} \label{eq:3}
    \mathbf{y}_z=\sqrt{\rho}\mathbf{H} x_z+ \mathbf{n}_z,
\end{equation}
where $\textbf{H} \in \mathbb{C}^{M \times K}$ is the channel matrix between the user and the BS; and  $\mathbf{n}_k \sim \mathcal{CN} (\mathbf{0},\sigma^{2}\mathbf{I}_M)$ represents the additive white Gaussian noise (AWGN), where each element follows a circularly symmetric complex Gaussian variable with zero mean and variance $\sigma^2$. Here, $\mathcal{CN}(\cdot,\cdot)$ denotes the circular symmetric complex Gaussian noise and  $\rho$ is the transmitted power allocated to each modulated data symbol. The signal-to-noise ratio (SNR) is be defined by
 $\mathrm{SNR} =\rho/\sigma^2$.
Once the BS has received all the noisy signal $\{ \mathbf{y}_z \}$, it  will decode and exploit the inverse projection to reconstruct the image, denoted by $\tilde{\mathbf{I}}$ as 
 $    \{ \mathbf{y}_z \} \overset{\mathcal{M}^{-1}}{\longmapsto} \mathbf{\tilde{I}},$  
where $\mathcal{M}^{-1}$ represents the demodulation or the inverse mapping. Let us simply denote $\mathbf{\tilde{I}} = \mathcal{M}^{-1}(\{ \mathbf{y}_z \})$. We emphasize that, from the signal received in \eqref{eq:3} and the given instantaneous channels, the BS decodes the image data information by the recovery model presented hereafter.
\vspace{-0.2cm}
\subsection{Problem Formulation}
\vspace{-0.1cm}
The received measurement data at the BS contains noise,  artifacts, and loss from the environment. The contamination  may be extremely severe, especially at a low SNR regime.  Our goal is to reconstruct the image $\tilde{\mathbf{I}}$ with the highest amount of desired information retained as the original semantic image data  $\mathbf{I}$. Consequently, the optimization problem, which we would like to solve, is formulated as follows 
\begin{equation} \label{eq:6}
\begin{aligned}
    && \underset{\tilde{\mathbf{I}}}{\mathrm{minimize}} & \quad \|\mathcal{M}^{-1}(\{ \mathbf{y}_z \}) - \mathbf{I} \|_{F}^{2}\\
    && \text{subject to} & \quad  \tilde{\mathbf{I}} = \mathcal{R}( \mathcal{M}^{-1}(\{ \mathbf{y}_z \}) ),\\
    &&& \quad x_z \in \mathcal{O}, \forall z,
\end{aligned}
\end{equation}
where $\| \cdot \|_F$ is the Frobenius norm.
The objective function of problem \eqref{eq:6} is to recover the semantic image data that is close to the original one. The mathematical operation $\mathcal{R}(\cdot)$ in the first constraint reconstructs the entire image from the partial image information. This optimization problem is non-convex from the mathematical points of view. Besides, the received measurements  are contaminated by the fading channels during transmission and noise at the receiver, so the useful information may be lost. In addition, problem~\eqref{eq:6} is only based on the features of the physical layer. In other words, it does not consider the image structure as a constraint, and therefore there is room to acquire a better the reconstruction quality than its locally optimal solution.
\vspace{-0.2cm}
\section{Image Recovery and Restoration}
\vspace{-0.1cm}
This section considers the two-stage image reconstruction process. In the first stage, the ADMM algorithm is brought to obtain an initial decoded image. Then, in the second stage, the image structure is exploited by a low-pass filter to enhance the reconstruction quality.
\vspace{-0.2cm}
\subsection{Image Recovery}
\vspace{-0.1cm}
By utilizing the ADMM algorithm in \cite{chinh2021performance,van2017block}, the first stage can obtain a local solution to problem~\eqref{eq:6}. This algorithm actually approximates the inherent nonconvexity of \eqref{eq:6} to a convex optimization problem. From the received measurement data $\{\mathbf{y}_z\}$ and the channel matrix $\mathbf{H}$,  the ADMM algorithm will iteratively decode $\tilde{x}_z$ which is the constellation point contained in the received data measurement. As wished, $\tilde{x}_z$ should be coincided with the modulated symbol $x_z$ sent by the user. From the set $\{ \tilde{x}_z\}$, one can attain the decoded image information, which is reshaped and restored in $\tilde{\mathbf{I}}$. As presented in the previous section, the decoded image may still not good in terms of, for example the BER, PSNR, and SSIM due to the locality of the solution. Therefore, after the first stage, the decoded image will be further processed to improve the image quality by expecting that texture and fine details of the image will be better captured. It is motivated by exploiting the image structure from the fact that natural images are strongly correlated. The correlation is indeed deployed in the lowpass filters and we can borrow them in our applications.
\vspace{-0.2cm}
\subsection{Image Restoration}
\vspace{-0.1cm}
In the second stage, the noise and residual errors are removed by a filtering process with an effective kernel,  which may be classified into two types: pixel-based kernel, e.g., Gaussian filter and median filter, and patch-based kernel, e.g., block-matching and 3D (BM3D) filter. This section will look at the basis of these techniques.

\subsubsection{Median filter}
is a non-linear filter, popularly used in image restoration due to its effectiveness at removing noise and preserving the strong edges. This filter is particularly effective against a typical noise called ``salt and pepper" noise. Specifically, the median is calculated by sorting all the pixel values of an image area (surrounding neighborhood) into numerical order, and then the median pixel value is utilized instead of the considered pixels.
\subsubsection{Gaussian filter}
is utilized to blur images and therefore remove noise. This filter works by using the bivariate distribution as a point-spead function. The goal is achieved by convolving an noisy image with the a Gaussian generated kernel. Mathematically, the filtered image is obtained as
\begin{equation}
    \hat{\mathbf{I}} = \tilde{\mathbf{I}} \circledast \mathbf{G},
\end{equation}
where $\circledast$ is the convolution operator; and $\mathbf{G} \in \mathbb{R}^{ u \times u}$ is the Gaussian kernel with $u$ being the kernel's size; and $\tilde{\mathbf{I}}$ is the noisy image. The  Gaussian kernel $\mathbf{G}$ is defined as
\begin{equation}
    \mathbf{G}(x,y)= \frac{1}{2\pi\sigma^2}e^{-\frac{x^2+y^2}{2\sigma^2}},
\end{equation}
where each pair $(x; y)$ stands for a coordinate in the kernel with $-u \leq x; y \leq u$ and $\sigma$ is the standard derivation of the distribution. We stress that the smoothness level of a Gaussian filter is controlled by changing the variance $\sigma$.
\subsubsection{BM3D filter}
is the patch-based noise reduction filter proposed by Dabov et. al. in \cite{2007-Image-BM3D-filtering}. The patch-based filtering algorithms have become popular in image processing nowadays. In the ideology of  noise reduction, one averages patches with the similar structure, noise will be cancelling each, particularly effective as noise is symmetric with zero mean. The advantage of this method is that it smooths out flat areas while preserving fine details and sharp edges. However, this type of filter often has high computational complexity due to the nonlocality of similar pixels, resulting in longer time consumption and the  penalty parameters are also nontrivial to select.

The proposed algorithm in \cite{2007-Image-BM3D-filtering} was based on transformed sparse representation in the frequency domain. The noisy image is first processed by extracting reference blocks from it and for each block, find blocks that are similar to it by using block matching and stack them together to form a three dimensional array (group). Then the collaborative filtering is performed of the group and the 2D estimates of all grouped blocks are returned to their original location. These estimated blocks can be overlap to each other, thus in the end, the authors aggregate all the estimates to form the final filtered imaged.
The advantage of this filter is that it both helps reduce the noise of the image while also preserving the details of the original image. 
\begin{remark}
This paper has focused on enhancing the reconstruction quality of images transmitted over the fading channels. The semantic information, e.g., the spatial correlation and noise information, among pixels are exploited to mitigate the remaining noise and artifacts. 
\end{remark}
\vspace{-0.2cm}
\section{Evaluation Metrics}
\label{Sec:EvaluationMetrics}
\vspace{-0.1cm}
In this section, we take a look at the metrics used to assess the performance of a telecommunications system and evaluate the quality of the image. Thereby, we have an intuitive view of the transmission efficiency and the influence of post-filtering process on the performance of a wireless network. In addition, we analyze the correlation of evaluation metrics used for wireless systems and for image signals.
\vspace{-0.2cm}
\subsection{BER Metric}
\vspace{-0.1cm}
When it comes to transmitting data from one point to another, the key metric to evaluate is how many errors will appear in the data that appears at the remote end, thus make the BER metric an important metric in measuring the performance of a communication system or a data channel. As the name implies, the BER is defined as the rate at which errors occur in a transmission system. The definition can be translated to the following formula
\begin{equation}
    \mathrm{BER} = E/B,
\end{equation}
where $E$ and $B$ is the total number of error bits and the total number of bits that are transmitted, respectively. Based on BER, we can evaluate the quality of the transmission process as well as the effectiveness of using digital filters in restoring the original image. In a communication system, the receiver side BER may be affected by different transmission factors, for instance, channel noise, interference, attenuation, multipath fading, etc.

BER is an metric that are mainly used at the physical layer, so it only considers the accuracy of the transmitted data to the original data without considering the actual quality of the image and the effect of noise. Therefore, we need to look at other metrics used to evaluate image quality. 
\begin{figure*}[t]
	\centering
	\begin{subfigure}{0.19\linewidth}
		\centering  
		\includegraphics[width=\textwidth]{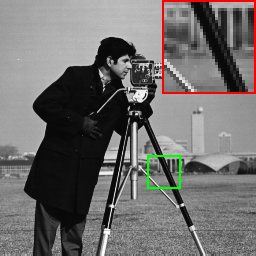}  
		\caption{Ground truth}
		\label{fig:sub-1}
	\end{subfigure}
	\begin{subfigure}{0.19\linewidth}
		\centering
		\includegraphics[width=\textwidth]{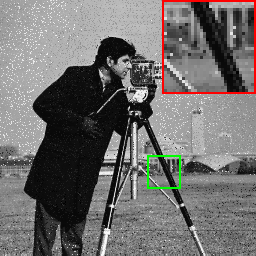}  
		\caption{Decoded image}
		\label{fig:sub-2}
	\end{subfigure}
	\begin{subfigure}{0.19\linewidth}
		\centering
		\includegraphics[width=\textwidth]{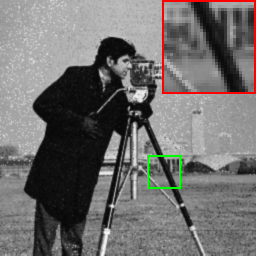}  
		\caption{Gaussian filter}
		\label{fig:sub-3}
	\end{subfigure}
	\begin{subfigure}{0.19\linewidth}
		\centering
		\includegraphics[width=\textwidth]{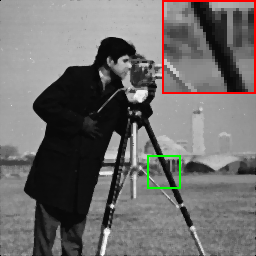}  
		\caption{Median filter}
		\label{fig:sub-4}
	\end{subfigure}
	\begin{subfigure}{0.19\linewidth}
		\centering
		\includegraphics[width=\textwidth]{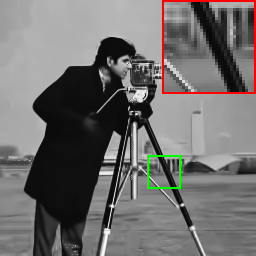}  
		\caption{BM3D filter}
		\label{fig:sub-5}
	\end{subfigure}
	\newline
	\begin{subfigure}{0.19\linewidth}
		\centering  
		\includegraphics[width=\textwidth]{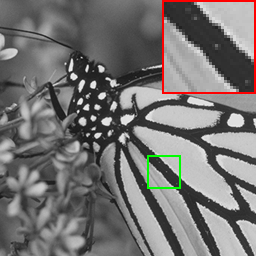}  
		\caption{Ground truth}
		\label{fig:sub-6}
	\end{subfigure}
	\begin{subfigure}{0.19\linewidth}
		\centering
		\includegraphics[width=\textwidth]{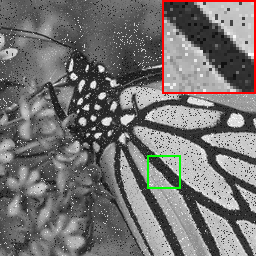}  
		\caption{Decoded image}
		\label{fig:sub-7}
	\end{subfigure}
	\begin{subfigure}{0.19\linewidth}
		\centering
		\includegraphics[width=\textwidth]{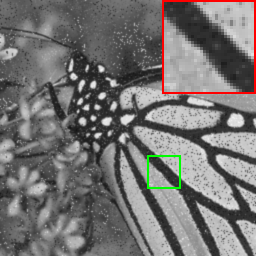}  
		\caption{Gaussian filter}
		\label{fig:sub-8}
	\end{subfigure}
	\begin{subfigure}{0.19\linewidth}
		\centering
		\includegraphics[width=\textwidth]{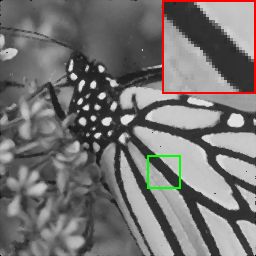}  
		\caption{Median filter}
		\label{fig:sub-9}
	\end{subfigure}
	\begin{subfigure}{0.19\linewidth}
		\centering
		\includegraphics[width=\textwidth]{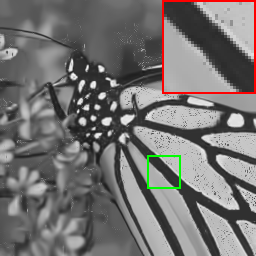}  
		\caption{BM3D filter}
		\label{fig:sub-10}
	\end{subfigure}
	\caption{Visual comparison by using the Cameraman and Monarch images from Set12 with the SNR equal to 10~[dB]: $(a)$ and $(f)$ are the ground truth; $(b)$ Decoded Cameraman image: BER = 0.0052, PSNR= 24.84 [dB] and SSIM = 0.43; $(g)$ decoded Monarch image:  BER=0.0054, PSNR = 21.22 [dB], and SSIM = 0.49; $(c)$ Post-filtered Cameraman image with Gaussian filter: BER=0.0375, PSNR=27.28 [dB], SSIM=0.48; $(h)$ Post-filtered Monarch image with Gaussian filter: BER=0.0391, PSNR=25.70 [dB], SSIM=0.63; $(d)$ Post-filtered Cameraman image with median filter: BER=0.0375, PSNR=26.69~[dB], SSIM=0.54; $(i)$ Post-filtered Monarch image with median filter: BER=0.0266, PSNR=29.11~[dB], SSIM=0.83; $(e)$ Post-filtered Cameraman image with BM3D filter: BER=0.0393, PSNR=29.51~[dB], SSIM=0.42; and $(j)$ Post-filtered Monarch image with BM3D filter: , BER=0.0403, PSNR=27.20~[dB], SSIM=0.69.}
	\label{fig:Sample10}
	\vspace*{-0.5cm}
\end{figure*}
\vspace{-0.1cm}
\subsection{PSNR Metric}
\vspace{-0.1cm}
The PSNR is an expression for the ratio of the maximum possible value (power) of a signal to the power of distortion noise that affects the quality of the signal. Because many signals have a very wide dynamic range, (ratio between the largest and smallest possible values of a changeable quantity) the PSNR is usually expressed in terms of the logarithmic decibel scale.

Assume that the data we are considering is a 2D digital image signal. Let $f$ be the original image data matrix of size size $m\times n$, $g$ is the image data matrix to be compared and must have the same size $m\times n$ as the original. Mathematically, PSNR is defined as follows:
\begin{equation}
    \text{PSNR} = 10 \text{log}_{10}\bigg(\frac{\mathrm{MAX}^{2}}{\mathrm{MSE}}\bigg)=20 \text{log}_{10}\bigg(\frac{\mathrm{MAX}}{\sqrt{\text{MSE}}}\bigg),
\end{equation}
where MAX is the maximum pixel value, depending on the original image data. For example, if the input image has a double-precision floating-point data representation, then MAX is 1. If it has an 8-bit representation, MAX is 255. The remaining parameter is MSE, which means the Mean Squared Error of the two image and is calculated as follows:
\begin{equation}
    \text{MSE} = \frac{1}{mn}\sum_{i=1}^{m}\sum_{j=1}^{n}\big[f(i,j-g(i,j)\big]^2, 
\end{equation}
MSE and the PSNR are both utilized to measure the quality of two images. The MSE represents the cumulative squared error between the compressed and the original image, whereas PSNR represents a measure of the peak error. The lower the value of MSE, the lower the error, and the higher the PSNR, the more similar restored image to the original image, thus means the better result acquired.  
The main limitation of this metric is that two degraded images with the same PSNR value could contain very different types of errors, some of which are easier to detect than others, resulting in different image quality.
\begin{figure*}
	\begin{minipage}{0.48\textwidth}
		\centering
		\includegraphics[trim=0.5cm 0cm 0.5cm 0.1cm, clip=true, width=3.2in]{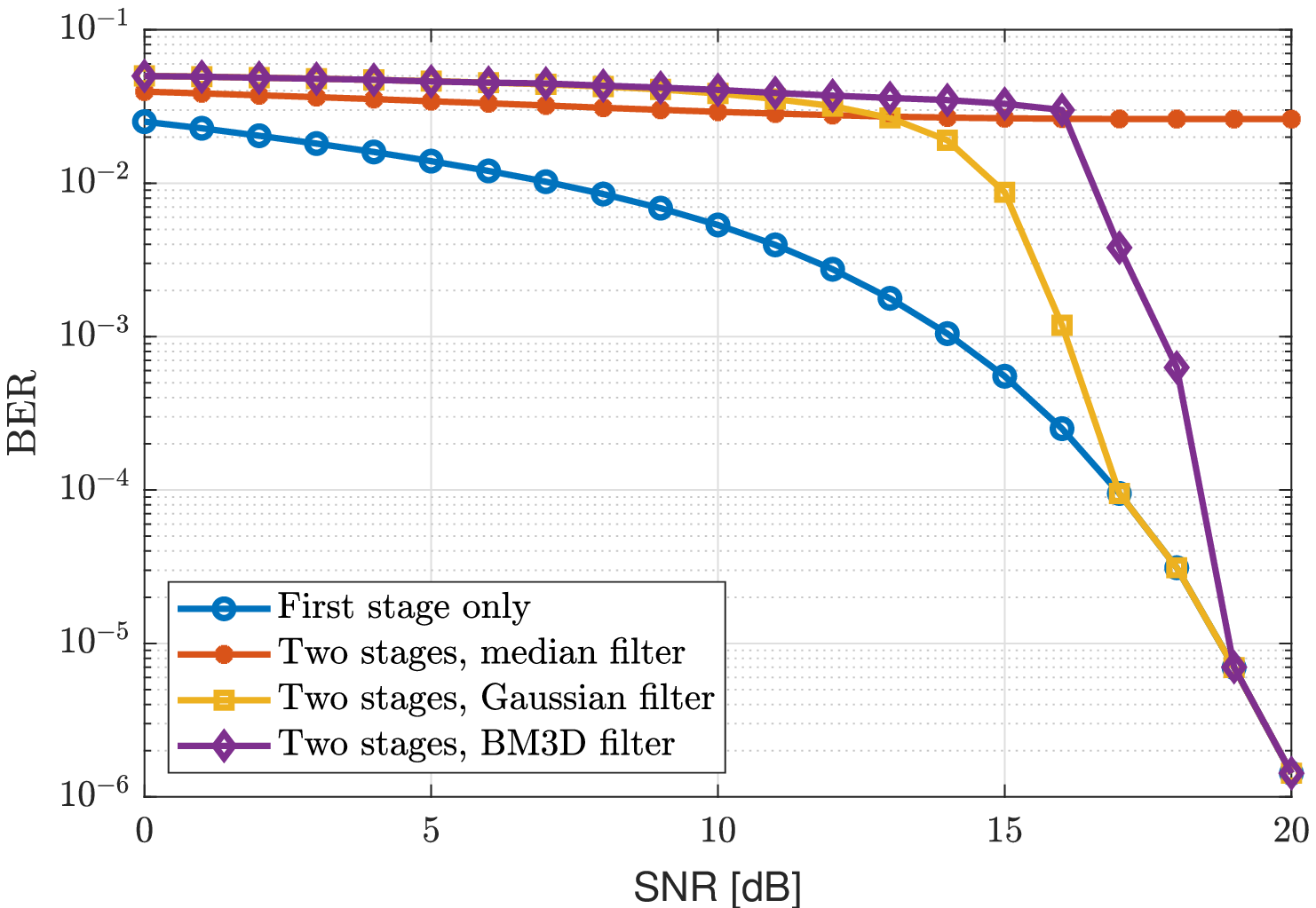}
		\caption{BER versus the different SNR [dB] values.}
		\label{fig:BER}
		\vspace*{-0.25cm}
	\end{minipage}
	\hfill
	\begin{minipage}{0.48\textwidth}
		\centering
		\includegraphics[trim=0.5cm 0cm 0.5cm 0.1cm, clip=true, width=3.2in]{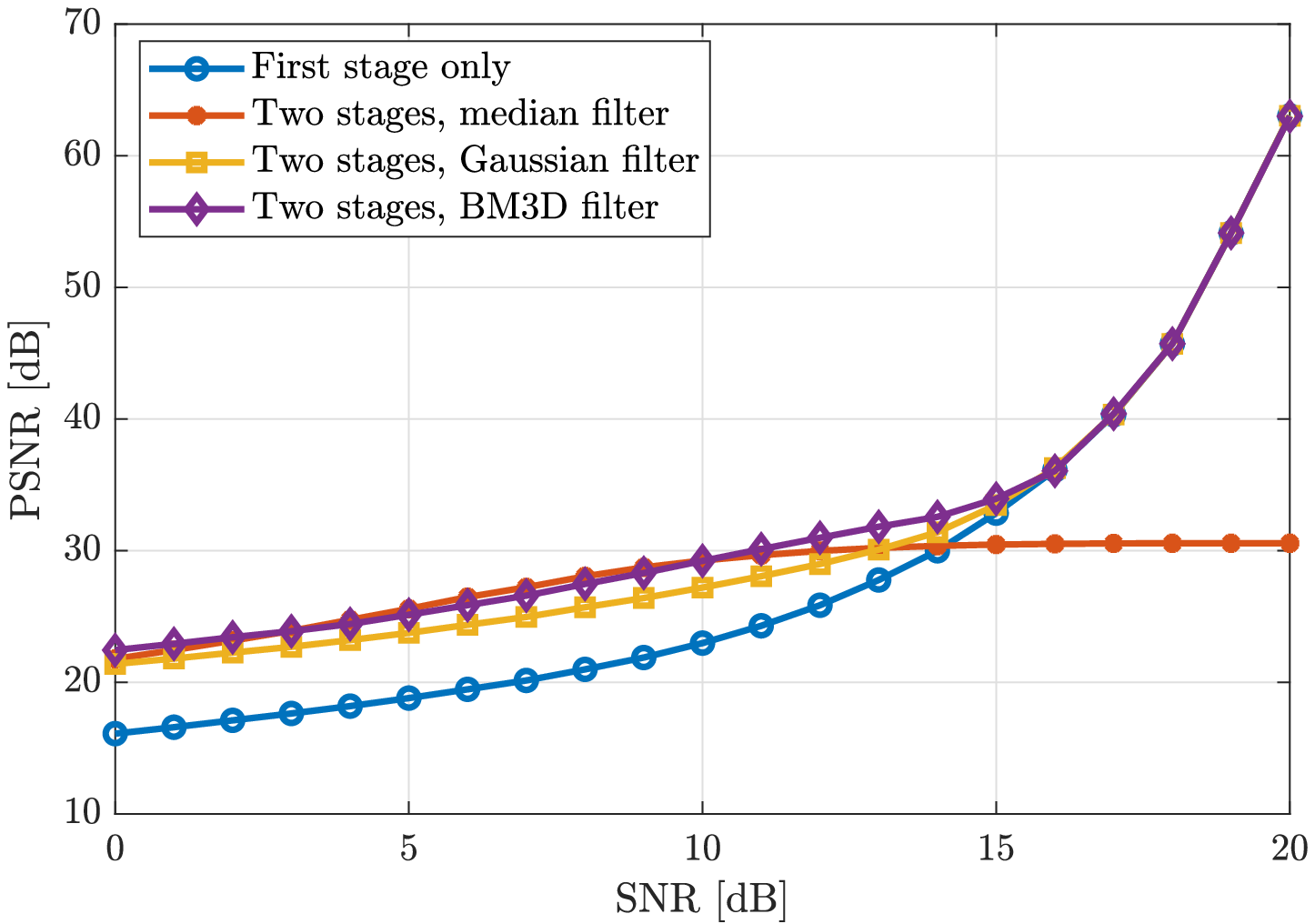}
		\caption{PSNR [dB] versus the different SNR [dB] values.}
		\label{fig:PSNR}
		\vspace*{-0.25cm}
	\end{minipage}
	\vspace*{-0.25cm}
\end{figure*}
\begin{figure}[t]
	\centering
	\includegraphics[trim=0.5cm 0cm 0.5cm 0.1cm, clip=true, width=3.2in]{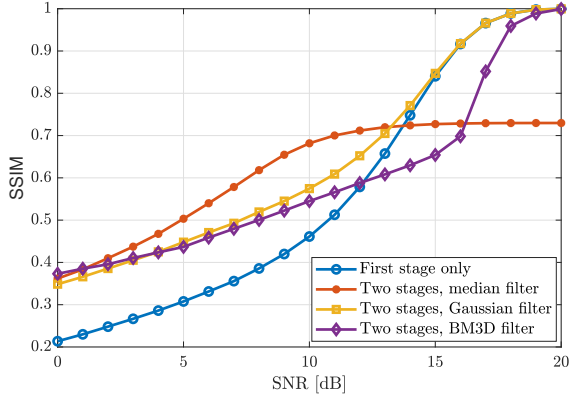}
	\caption{SSIM versus the different SNR [dB] values.}
	\label{fig:SSIM}
	\vspace*{-0.5cm}
\end{figure}
\vspace*{-0.5cm}
\subsection{SSIM Metric}
 Unlike the PSNR, the SSIM metric is often utilized to evaluate the similarity or similarity between the noisy signal and the original data. It was first proposed in \cite{2004-SSIM} by Zhou Wang, based on the assumption that human visual perception is highly adaptive to copying structural information from an image. This assessment is designed by modeling any image data using three factors: luminance, contrast and structure \cite{2010-PSNR-vs-SSim}.\\
To calculate the SSIM value of two image data f and g, first we must calculate the illuminance value of each image:
\begin{equation}
    \mu_x=\frac{1}{N}\sum_{i=1}^{N}x_i,
\end{equation}
where $N$ is the number of pixels in the image $\mathbf{x}$, i.e., $\mathbf{x}=[x_1, \cdots, x_i, \cdots, x_N]^T$, thus making illuminance the average of all the pixels. After measuring the illuminance, we calculate the contrast which is the standard deviation of all pixels as follows
\begin{equation}
    \sigma_x=\bigg[\frac{1}{N-1}\sum_{i=1}^{N}(x_i - \mu_x)^2\bigg]^{\frac{1}{2}}.
\end{equation}
The SSIM formula is based on three comparison measurements between the input image: luminance, contrast and structure. However, from these two values, we can calculate the SSIM index based on the reduced formula as follows
\begin{equation}
    \mathrm{SSIM}(f,g) = \frac{(2\mu_f \mu_g +C_1)(2\sigma_{\textit{fg}}+C_2)}{(\mu_f^2+\mu_g^2+C_1)(\sigma_f^2+\sigma_g^2+C_2)},
\end{equation}
in which $C_1=(K_1L)^2$ , $C_2=(K_2L)^2$ are the constants with \emph{L} being the dynamic range of pixel values and the constant $K1, K2\ll 1$. The covariance between two images $f$ and $g$ (reshaped into vectors of the $N$ elements), denoted by $\sigma_{fg}$, is defined as
\begin{equation}
    \sigma_{fg}=\frac{1}{N-1}\sum_{i=1}^{N}(f_i - \mu_f)(g_i-\mu_g).
\end{equation}
The resultant SSIM index is a decimal value between 0 and 1, and value 1 is only reachable in the case of two identical sets of data and therefore indicates perfect structural similarity. A value of 0 indicates no structural similarity. The difference with other techniques such as MSE or PSNR is that structural information considers pixels to have strong inter-dependencies with each other, especially when they are spatially close. These dependencies carry important information about the structure of the objects in the visual scene. 

\section{Experimental Evaluation}
This section provides  the simulation results of the natural image data recovery process for the uplink data transmission of a massive MIMO communication system. By varying the SNR values, numerical results evaluate the reconstruction quality of the system by utilizing the metrics BER, PSNR, SSIM as presented in Section~\ref{Sec:EvaluationMetrics}. The base station is equipped with $64$ antennas, while the user terminal has 4 antennas. We use  $256$-QAM to modulate and map image data onto the corresponding constellation points before transmitting over the wireless channels. In order to scrutinize the communication reliability, the SNR value is considered in the range from $0$~[dB] to $20$~[dB]. The filtering parameters are selected to attain the best system performance. In particular, at the SNR$=0$~dB, the standard derivation of the Gaussian filter is $1.3$ and that of the BM3D filter is $41$.  the filtering parameter is decreased the SNR gets higher due to less residual errors. The propagation channels follow the Rayleigh fading distribution and  noise follows a circular symmetric complex Gaussian distribution.
\vspace{-0.2cm}
\subsection{Visualization Comparison}
\vspace{-0.1cm}
We observe the representation of reconstructed images with the SNR value of 10~[dB] in Fig.~ \ref{fig:Sample10}. The images in the first column are the original images or called the ground truth. The second column includes the decoded images after applying the ADMM algorithm. Meanwhile, the refined images result from deploying the Gaussian, median, and BM3D filters, respectively. We can observe that the decoded image by the ADMM algorithm still contains lots of noise and artifacts. This is because the ADMM algorithm considered in this paper decodes the image data based on the channel information only. By applying a low-pass filter to smooth out the decoded image, we observe much the better restoration quality. Noise and artifacts have been greatly reduced by using the Gaussian and median filters, but besides that, the edges and details are also lost. More advance, the BM3D filter offer the ability to preserve the edges and details of the image, but cannot completely remove the noise from the image, especially the noise in the Monarch image is still clearly visible. Median filters are very good at handling salt and pepper noise type due to their characteristics. Details are retained relatively well, comparable to the results of the BM3D filter, like the tripod in the Cameraman image. In the following, we investigate the system performances by different metrics as mentioned.
\subsection{BER Performance}
In Fig.~\ref{fig:BER}, we visualize the simulation results of the average BER value of 12 natural images in the Set12 dataset \cite{Zhang2017residual}.
It shows that the blue line corresponding to the case where the image is restored without a filter gives the lowest bit error rate and gradually decreases inversely with the SNR of the channel, means that when the channel's state gets better, the BER would decrease. However, as the SNR increases, the filtered image's BER stays very high and does not seem to improve, which leads to the conclusion, that the filters are not able to restore the recovered image to its original version. Instead, it will sacrifice image details to increase the quality, as can be seen in Fig. \ref{fig:Sample10}.
This seems to be contradictory because in the previous section, we noticed that utilizing filters improve the image quality significantly compared to contaminated image after transmission and restoration using the ADMM algorithm. To clarify this disagreement, we need to understand that the purpose of ADMM algorithm is to solve the optimization problem and minimize the transmission error. Therefore, any subsequent action by the filter would increase the difference of data bits, resulting in a higher BER. This is the reason why the average BER value in the case of using the filter is not as good as the BER value in the case of not using the filter. It leads to the conclusion: For structured data such as digital image signals, BER is no longer a decisive criterion for evaluating the quality of the recovered data. 
\subsection{PSNR Performance}
As mentioned above, the BER index in this case is not suitable for assessing the quality of the recovered image. Therefore, we need to use metrics commonly used in image processing. Figure \ref{fig:PSNR} depicts the average PSNR value results of the different cases mentioned earlier. The results show that with the PSNR index, the image quality after going through the filter is significantly higher than the image at the output of the ADMM algorithm. The Median filter and the BM3D filter give approximately the same results, but the weak point of the Median filter is that it does not have a control parameter, leading to a dropped image quality at higher channel's SNR.
\subsection{SSIM Performance}
Another parameter commonly used in image processing to evaluate quality is the SSIM index. Fig.~\ref{fig:SSIM} is the average SSIM result of the images in the Set12 dataset. In general, the image after passing through the filter gives significantly better SSIM results than the original image, especially in the low and medium SNR regions. The Median filter performs best in this area but as the channel quality gets better, similar to PSNR, the output image no longer improve. The BM3D filter here due to being optimized for the highest PSNR result did not give good performance at SSIM metric.

\section{Conclusion}
This paper investigated the content-based image transmission over the wireless networks. We focused on the image recovery  at the BS with the two-stage reconstruction process. First, the BS exploits the channel state information to decode the image data from the received noisy measurements. After that, the semantic information, such as spatial correlation, is explicitly exploited to refine the reconstructed image. We have demonstrated by numerical results that the image content can be effectively manipulated to refine the reconstructed images. Moreover, the traditional evaluation metrics may be conflicted to each other. Consequently, defining a new evaluation metric to cover multiple features of semantic communications should be a potential research direction for future work.
\bibliography{citation.bib}
\bibliographystyle{IEEEtran}

\end{document}